\documentclass[%
 reprint,
 amsmath,amssymb,
 aps,
]{revtex4-2}
\usepackage{mathptmx}
\usepackage{graphicx}
\usepackage{dcolumn}
\usepackage{bm}
\usepackage[colorlinks=true
,urlcolor=blue
,anchorcolor=blue
,citecolor=blue
,filecolor=blue
,linkcolor=blue
,menucolor=blue
,pagecolor=blue
,linktocpage=true
,pdfproducer=medialab
,pdfa=true
]{hyperref}

\usepackage{amsmath}
\usepackage{hyperref}
\usepackage{physics}
\usepackage{graphicx}
\usepackage{xcolor}
\usepackage{MnSymbol}
\usepackage{float}
\usepackage{relsize}
\usepackage{romannum}
\newcommand{\beq}{\begin{eqnarray}}
\newcommand{\eeq}{\end{eqnarray}}
\def\l{\left(}
\def\r{\right)}
\def\nn{\nonumber}
\def\la{\mathcal{L}}

\def\={&=&}
\def\p{\partial}

\newcommand{\bwt}{\begin{widetext}}
\newcommand{\ewt}{\end{widetext}}
\def\nn{\nonumber}

\begin{document}

\preprint{APS/123-QED}

\title{Revisiting Vacuum Energy in Compact Spacetimes}

\author{S R Haridev}
\email{p20180460@hyderabad.bits-pilani.ac.in}
\author{Prasant Samantray}%
 \email{prasant.samantray@hyderabad.bits-pilani.ac.in}
\affiliation{Department of Physics, BITS- Pilani Hyderabad Campus, Jawahar Nagar, Shamirpet Mandal, Secunderabad 500078, India}

\date{\today}

\begin{abstract}
We revisit the calculation of vacuum energy density in compact spacetimes. For the general case of $k$ compact spatial dimensions in $p+k$ dimensional Minkowski spacetime, we calculate the Casimir force on a piston placed in the presence of external electromagnetic fields. We observe that while the presence of a strong external magnetic field reduces the intensity of the Casimir force, the presence of an electric field enhances it instead. We discuss various physical cases, and also derive the particle production rate in the presence of such external fields in compact spacetimes. We observe that the pair production rate is enhanced in the presence of a piston.
\end{abstract}

\keywords{Casimir Force, Vacuum energy, Higher dimensional theory }
\maketitle
\flushbottom

\section{Introduction}\label{sec:intro}
Almost a century ago Kaluza unified general relativity with electromagnetism by invoking a fifth extra compact spatial dimension \cite{kaluza1921sitzungsber,klein1991quantum}. While the idea was profound, it did not exactly take off as a fundamental framework of our universe. Decades later superstring theory - the most promising candidate for unifying all the known forces at present, required the existence of compact extra spatial dimensions to mandate the theory to be mathematically consistent. From a phenomenological standpoint these string theories come tantalizingly close to explaining the particle content of our observable universe. Well almost. This naturally renewed interest in spacetimes with extra spatial dimensions. For example, also see \cite{klein1991quantum,kaluza2018unification}.

Tangentially, in standard quantum field theory in Minkowski spacetime, the zero point energy is infinite and is usually subtracted out by shifting the datum of the Hamiltonian. This is done via the prescription of normal ordering and the infinite zero point energy or the vacuum energy has no physical meaning. However, in the presence of compact extra spatial dimensions, this zero point/vacuum energy has non-trivial meaning and in principle can have observational signature like the force on a plate, or on a pair of plates placed at a distance etc. This inevitably ties the physics of compact dimensions with the well known Casimir effect. Along with the Unruh effect and the Schwinger effect, the Casimir effect is one of the cornerstones of standard quantum field theory which has been experimentally verified. The result of this effect is that if two parallel plates are kept at a distance, they attract each other by exerting a force known famously as the Casimir force \cite{casimir}. Casimir effect/force is the result of introducing a length scale/boundary condition on the quantum field which otherwise is generally absent. Since compactification introduces a natural length scale, it is expected that this will naturally lead to a non-trivial zero point/vacuum energy even in the absence of any other additional boundary conditions. In fact, there has been a host of literature on these issues \cite{cheng1,cheng2,frank2007casimir,
poppenhaeger2004casimir,linares2008casimir,perivolaropoulos2008vacuum,pascoal2008estimate,
edery2008compact,teo2009finite,elizalde2009repulsive,saharian2008casimir,Alvarez_2021}. Additionally placing a pair of plates in such a compactified spacetime leads to some interesting results and the reader is urged to refer to \cite{accetta1986finite,
goldberger2000quantum,nam2000casimir,huang2001finite,ponton2001casimir,
hofmann2001stabilization,huang2001casimir,graham2002calculating,elizalde2003casimir} for more details. In particular, \cite{cheng1} considered the Casimir force on a plate placed in Minkowski space with extra compactified dimension and calculated a seemingly counter-intuitive repulsive Casimir force in the asymptotic limit. The same author a couple of years later considered a piston but this time in a finite box in the compact spacetime, and concluded that the repulsive force was neutralized in the asymptotic limit \cite{cheng2}. In \cite{poppenhaeger2004casimir} the authors considered the extra dimension with electromagnetic field and computed an analytic expression for Casimir force. They also compared with the experimental data available and placed upper bounds on the size of extra dimensions. In fact, Casimir effect has also been studied in the context of stabilizing the extra compact dimensions, early universe cosmology and also in the context of string theory. Further commentaries and investigations on all these issues and more can be looked up in 
\cite{elizalde2001matching,gardner2002primordial,milton2002dark,setare2004stress,
mahajan2006casimir,doran2006measurable,fabinger2000casimir,gies2003casimir,hadasz2000casimir,fulling2, booktoms,Juric:2018qdi}. 

While most previous works on this issue have focused using the mode-sum approach and subsequent regularization procedures to extract the vacuum energy density and calculate Casimir force on plates. We in the present work use the method of effective action via the heat kernel method to do the same. This method allows for approximation schemes which allows us to calculate the vacuum energy in various different setting with a considerable degree of accuracy and ease as we shall see in later sections. We also use this formalism to calculate the Casimir force on a piston placed along the compact spatial dimension which is more direct and straightforward as compared with the standard mode-sum approach in the earlier approaches. We later generalize our approach to including electromagnetic fields, and also looking at particle production.

The organization of this paper is as follows. In section \ref{sec1} we discuss the mathematical tools used throughout the paper. In section \ref{seccon} we calculate the vacuum energy of a massive complex scalar field in $d$ dimensional spacetime with $k$ compact spatial dimensions. We use this result to calculate the Casimir force per unit area on a piston placed in the same manifold. In sections \ref{secb} and \ref{sece}, we extend these calculations to include the presence of a constant magnetic field perpendicular to the plates and an electric field parallel to the plates, respectively. We also calculate the Schwinger pair production in higher dimensions with and without the piston in such compact spacetimes. Finally, we summarize our results in section \ref{summary} and conclude with a discussion.

\section{General mathematical frame work}\label{sec1}
This section gives a general outline of the mathematical tools we used throughout the paper. Throughout this paper we works in $\hbar=c=1$ and our metric convention is $(-,+,+,+)$. Consider a complex scalar field of mass $m$ and charge $q$ coupled to a vector potential $A_{\mu}$. The Lagrangian density of the system is
\beq
\la \=-\l D_{\mu}\phi\r^{*}\l D^{\mu}\phi\r -m^{2}\phi^{2},
\eeq
where $D_{\mu} = \p_{\mu}-i q A_{\mu} $ is the gauge derivative. We work in Lorentz gauge ($\p_{\mu}A^{\mu}=0$) and in the Euclidean time ($\tau=i t$). The Euclidean effective potential $V_{E}$ for the theory is 
\beq\label{eupot}
V_{E} =\int_{0}^{\infty}\frac{K(s;x,x)}{s}ds,
\eeq
where $K(s;x,x)$ is the heat kernel $K(s;x,x')$ in the coinciding limit ($x\rightarrow x'$). Heat kernel satisfies the differential equation
\beq\label{heateqn}
\hat{D} K\l s; x,x'\r =- \frac{\p}{\p s}K\l s; x,x'\r,
\eeq
with boundary condition
\beq\label{bc1}
K(0; x,x')\= \delta(x-x'),
\eeq
and
\beq
\hat{D}\= -\p_{\mu}\p_{\mu}+2iqA_{\mu}\p_{\mu}+q^{2}A^{2}+m^{2}.
\eeq
We solved Eq.  (\ref{heateqn}) with boundary condition (\ref{bc1}) by expanding the heat kernel in Fourier basis as
\beq\label{heatfour}
K(s;x,x')\propto \sumint e^{i p x}\;\tilde{k}(s,p,...),
\eeq
where the summation is over the Fourier modes along the compact dimensions. Substituting the result of the heat kernel in Eq.  \ref{eupot} gives the Euclidean effective potential from which one can deduce the vacuum energy of the field and the corresponding Casimir pressure exerted on a parallel plate.
\section{Complex scalar field with constant gauge potential}\label{seccon}
To begin with, we consider a massive complex scalar field operating in $1+3$ spacetime dimension with one spatial dimension compactified ($x_{1}\sim x_{1}+L$). As the simplest case, we consider the case of periodic boundary conditions for the complex scalar field along the compact dimension. Since the topology is non-trivial, we can not gauge transform away the gauge potentials along the compact dimensions. So, we take the background gauge field to be $A_{\mu}=(0, a,0,0)$. For this gauge potential, we can solve Eq. (\ref{heateqn}) by writing the heat kernel in the Fourier modes (Eq. \ref{heatfour}), which gives the Euclidean effective potential as

\beq
V_{E}\= \frac{1}{8L\pi^{\frac{3}{2}}}\sum_{n=-\infty}^{\infty}\int_{0}^{\infty}\frac{ds}{s^{\frac{5}{2}}}e^{-M_{n}^{2}s},
\eeq
where $M_{n}^{2} = m^{2}+\l \frac{2\pi n}{L}-qa\r^{2}$. Using analytic continuation of gamma function
\beq\label{eupotnocom}
V_{E}\=\frac{4\pi^{2}}{3L^{4}}\sum_{n=-\infty}^{\infty}\l \l n- \frac{qaL}{2\pi}\r^{2}+\l\frac{mL}{2\pi}\r^{2}\r^{\frac{3}{2}}.
\eeq
We use the result from complex analysis for the infinite sum \cite{toms} which states
\beq\label{sum}
\sum_{n=-\infty}^{\infty}\left((n+\beta)^{2}+\alpha^{2}\right)^{-\lambda} \= \sqrt{\pi}\frac{\Gamma\left(\lambda-\frac{1}{2}\right)}{\Gamma(\lambda)}\alpha^{(1-2\lambda)}\nn\\
&&+4\sin(\lambda\pi)f_{\lambda}(\alpha,\beta),
\eeq
where
\beq\label{fff}
f_{\lambda}(\alpha,\beta) \= \mathcal{R}\left\{ \int_{\alpha}^{\infty}dx \frac{(x^{2}-\alpha^{2})^{-\lambda}}{e^{2\pi x+2\pi i\beta}-1}\right\}.
\eeq
Here $\mathcal{R}$ stands for the real part of the complex-valued function. The summation in Eq.  (\ref{sum}) is convergent for $\mathcal{R}(\lambda)>1/2$, but we have an analytic continuation of the same on the right side of the equation, which is valid for all the values of $\lambda$.  Using the result we can write Eq. (\ref{sum}) as
\beq
V_{E}\= \frac{m^{4}}{16\pi^{2}}\Gamma(-2)+\frac{16\pi^{2}}{3L^{4}}f_{-\frac{3}{2}}(\alpha,\beta),
\eeq
where $\alpha=(mL/2\pi)$ and $\beta= -(qaL/2\pi)$.  First term in the potential corresponds to the contribution from $1+3$ dimensional flat topology, which one discards by renormalizing the vacuum energy of $1+3$ dimensional Minkowski spacetime to zero. Then the minimized Lorentzian effective potential ($V=-V_{E}$) is (appendix \ref{appf})
\beq
V_{min} = -\frac{16\pi^{2}}{3L^{4}}f_{-\frac{3}{2}}(\alpha,0).
\eeq
Changing the integration variable in Eq. (\ref{fff}) to $y=x/\alpha$ and using the modified Bessel function of second kind $K_{\nu}(z)$ we can write
\beq\label{bessel}
f_{\lambda}(\alpha,0)\=\frac{\Gamma(1-\lambda)}{\pi^{1-\lambda}\alpha^{\lambda-\frac{1}{2}}}\sum_{n=1}^{\infty}n^{\lambda-\frac{1}{2}}K_{\lambda-\frac{1}{2}}(2\pi n \alpha).
\eeq
Similarly, one can calculate the vacuum energy of a complex scalar field operating in $d$ dimensional spacetime with $k$ compact spatial dimensions.  For $k$ compact spatial dimensions, we have $k$ summations in Eq.  \ref{eupotnocom} and each of the summation is converging (see appendix \ref{appf}). So we have $k!$ ways of doing this summation. Considering this fixes the symmetry in exchanging of compact dimensions ($L_{i}\leftrightarrow L_{j}$). 'sym' in the below equations corresponds to this symmetry fixing. Then the renormalized effective potential for a complex scalar field in $d$ dimensional spacetime with $k$ compact spatial dimension is

\beq\label{ren}
V_{ren}\=-\frac{4\pi\hbar}{\mathcal{V}_{k}}\sum_{s=1}^{k}\Bigg[\frac{(\pi)^{\frac{d-s}{2}}}{\Gamma\left(\frac{d-s}{2}+1\right)} \left(\frac{1}{L_{k+1-s}}\right)^{d-s}\left(\prod\limits_{r=1}^{k-s}L_{r}\right)\nn\\
&&\mathlarger{\mathlarger{\sideset{}{^*}\sum}}_{l}f_{_{\frac{s-d}{2}}}\left(\alpha_{k+1-s}^{k}, \beta_{k+1-s}\right)\Bigg]+\mbox{sym},
\eeq
where $\sideset{}{^*}\sum$ is a multiple summation and we have to sum over all $l\in (n_{k},...,n_{k+2-s})| s\ge 2$ in the range $(-\infty,\infty)$, $\mathcal{V}_{k}=L_{1}...L_{k}$ and 
\beq\label{alpha}
\alpha_{k+1-s}^{k} \= L_{k+1-s}\Bigg(L_{k+2-s}^{-2}\left(n_{k+2-s}+\beta_{2}\right)^{2}...\nn\\
&&+L_{k}^{-2}\left(n_{k}+\beta_{k}\right)^{2}+\l\frac{m}{2\pi}\r^{2}\Bigg)^{\frac{1}{2}},
\eeq
where $\beta_{k}=-qa_{k}L_{k}/2\pi$. To get Eq. (\ref{ren}) we used the mirror identity $\Gamma(-x)\sin(-\pi x)=\pi /x\Gamma(x)$. The renormalized effective potential obtains its minimum value when $a=0$ (see appendix \ref{appf}). Then using Eq.  (\ref{bessel}) the vacuum energy of a massive complex scalar field is
\beq\label{convmin}
V_{min}= -\frac{4\hbar}{L^{d}}\sum_{s=1}^{k}\sideset{}{^*}\sum_{l}&&\sum_{n=1}^{\infty}\l N_{k+1-s}^{k}\r^{\frac{d-s+1}{2}}n^{\frac{s-d-1}{2}}\nn\\
&&K_{\frac{d-s+1}{2}}(2\pi nN_{k+1-s}^{k})+\mbox{sym},
\eeq
where 
\beq\label{N}
N_{k+1-s}^{k}\= \sqrt{n_{k+2-1}^{2}+...+n_{k+1-s}^{2}+\l\frac{mL}{2\pi}\r^{2}}.
\eeq
In Eq. (\ref{N}) we assume $L_{1}\approx L_{2}...\approx L_{k}\approx L$. The function $z^{\lambda}K_{\lambda}(z)$ decays rapidly as $z$ varies from $0$ to $\infty$, this allows us to approximate the sum $\sideset{}{^*}\sum$ by its leading order as
\beq\label{apprx}
V_{min}\approx && -\frac{4\hbar}{\mathcal{V}_{k}}\sum_{s=1}^{k}\l \frac{\prod_{r=1}^{k-s}L_{r}}{L_{k+1-s}^{d-s}}\r\l\frac{mL_{k+1-s}}{2\pi}\r^{\frac{d-s+1}{2}}\times\nn\\
&&\sum_{n=1}^{\infty}n^{\frac{s-d-1}{2}}K_{\frac{d-s+1}{2}}(nmL_{k+1-s})+\mbox{sym}.
\eeq
This approximation is valid even for compact dimensions of difference size.  Also using the result $\lim_{z\rightarrow 0}z^{\nu}K_{\nu}(z)=2^{\nu-1}\Gamma(\nu)$ we can use Eq.  (\ref{apprx}) for calculating the vacuum energy of massless scalar field in the same spacetime. For the case of a masseless scalar field, this approximation gives a relative error of $1.3\%$, $3.9\%$ and $9\%$ for $\mathbb{R}^{(1, 1)} \times \mathbb{S}^{1}\times \mathbb{S}^{1}$, $\mathbb{R}^{(1, 3)} \times \mathbb{S}^{1} \times \mathbb{S}^{1}$ and $\mathbb{R}^{(1, 3)}\times\mathbb{S}^{1} \times \mathbb{S}^{1} \times \mathbb{S}^{1}$, respectively when we compare the result with Eq. (\ref{convmin}). The relative error increases as we increase the number of dimensions but we can consider higher order corrections in the same way. We use this result (Eq. \ref{apprx}) for calculating the Casimir force in the coming sections as we are interested in the qualitative change of Casimir force due to background fields and extra compact dimensions.
\subsection{Casimir force on a piston}\label{secforce}
\begin{figure}
\centering
\includegraphics[scale=0.4]{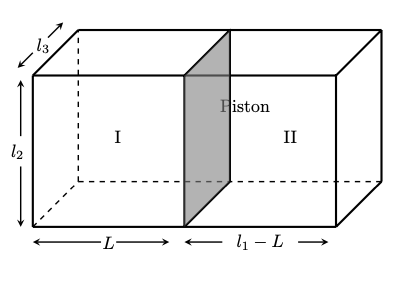}
\caption{A Rectangular piston}
\label{piston}
\end{figure}

\begin{figure}
\centering
\includegraphics[scale=0.6]{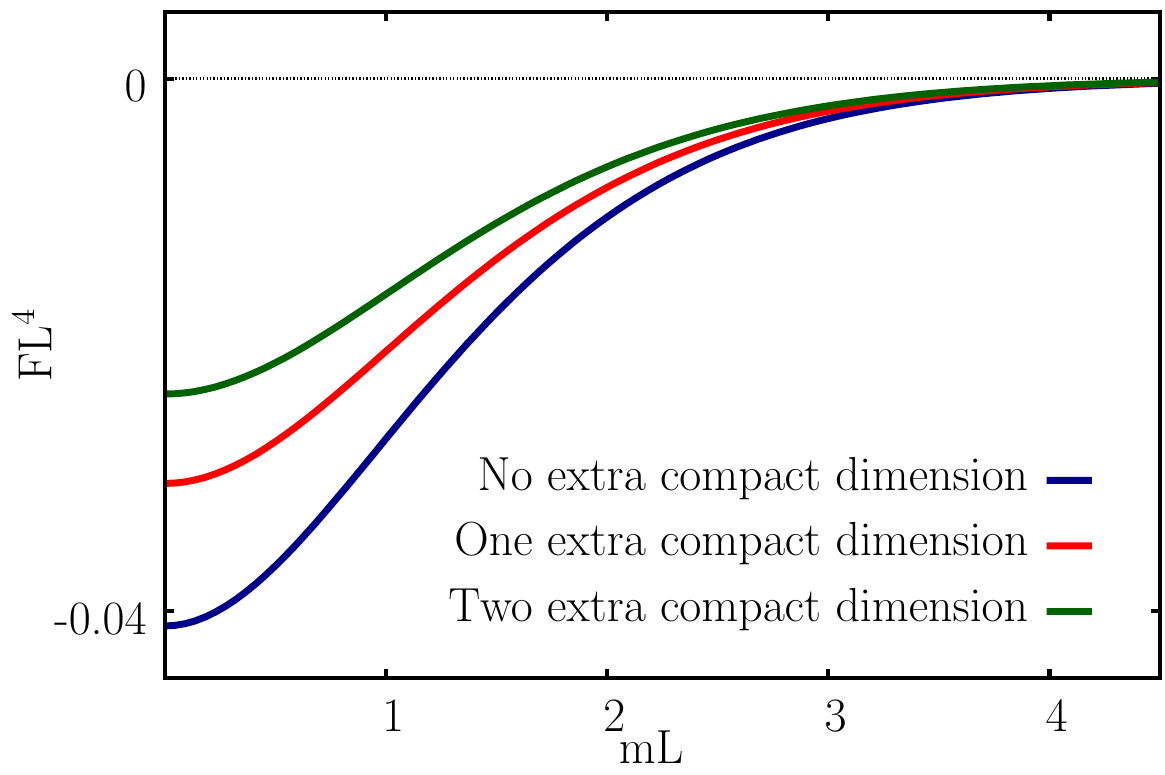}
\caption{Shows the variation of Casimir force with respect to $L m$.  In this plot we assume $R=L$. }
\label{figmass}
\end{figure}

Armed with our expression for $V_{min}$ (Eq. \ref{apprx}), we now turn our attention to calculate the Casimir force on a piston (or two parallel plates) placed in various compact spacetime dimensions. The energy density $\rho$ for a complex scalar field satisfying Dirichlet boundary condition (i.e., the field vanishing at boundaries $x=0$ and $x=L$) can also inferred from our results by replacing $L$ by $2L$ (where $L \in \{L_{1},...,L_{k}\}$)\cite{booktoms,pinto} as 
\beq\label{rho}
\rho(L)\= V_{min}(2L).
\eeq
 For calculating the Casimir force on a piston we consider the finite box boundary condition as shown in Fig. \ref{piston} and eventually take the limits $l_{1},l_{2},l_{3}\rightarrow \infty$. The importance of taking the finite box boundary condition is well discussed in \cite{cheng1,cheng2,fulling1,fulling2,piston}. 
The Casimir force has contributions from both regions \Romannum{1} and \Romannum{2} as
\beq\label{fforce}
\frac{F}{\mathcal{V}_{k}} \= -\frac{d}{dL}\left(L\rho_{1}\right)-\frac{d}{dL}\left((l_{1}-L)\rho_{2}\right)\Big|_{l_{1}\rightarrow\infty},
\eeq
where $\rho_{1}$ and $\rho_{2}$ are the energy densities in the region I and II respectively. Considering $d=4$ and $k=1$ in Eq. (\ref{apprx}) and using Eq. (\ref{rho}) and (\ref{fforce})
\beq\label{nocom}
F\= -\frac{m^{2}}{4L^{2}\pi^{2}}\sum_{n=1}^{\infty}\l \frac{2 m L}{n}K_{1}(\xi)+\frac{3}{n^{2}}K_{2}(\xi)\r,
\eeq
where $\xi = 2mnL$. This gives the force per unit area on the piston due to the vacuum fluctuations of a massive complex scalar field placed in a Minkowski spacetime. Eq.  (\ref{nocom}) is a known result in the literature \cite{pinto}, and in the limit, $m\rightarrow 0$ this gives the standard result first obtained by H B G Casimir \cite{casimir}. Similarly, we can calculate the force per unit area on the piston in the presence of extra compact dimensions (as shown in Fig. \ref{figmass}), and each extra compact dimension adds corrections to Eq. (\ref{nocom}). The correction vanishes in the limit $R\rightarrow 0$, where $R$ is the size of the extra compact dimension. Also, the limit $R\rightarrow\infty$ gives the results for $d$ dimensional flat topology \cite{massd}.
\section{Constant magnetic field perpendicular to the piston}\label{secb}
\begin{figure}[H]
\centering
\includegraphics[scale=0.7]{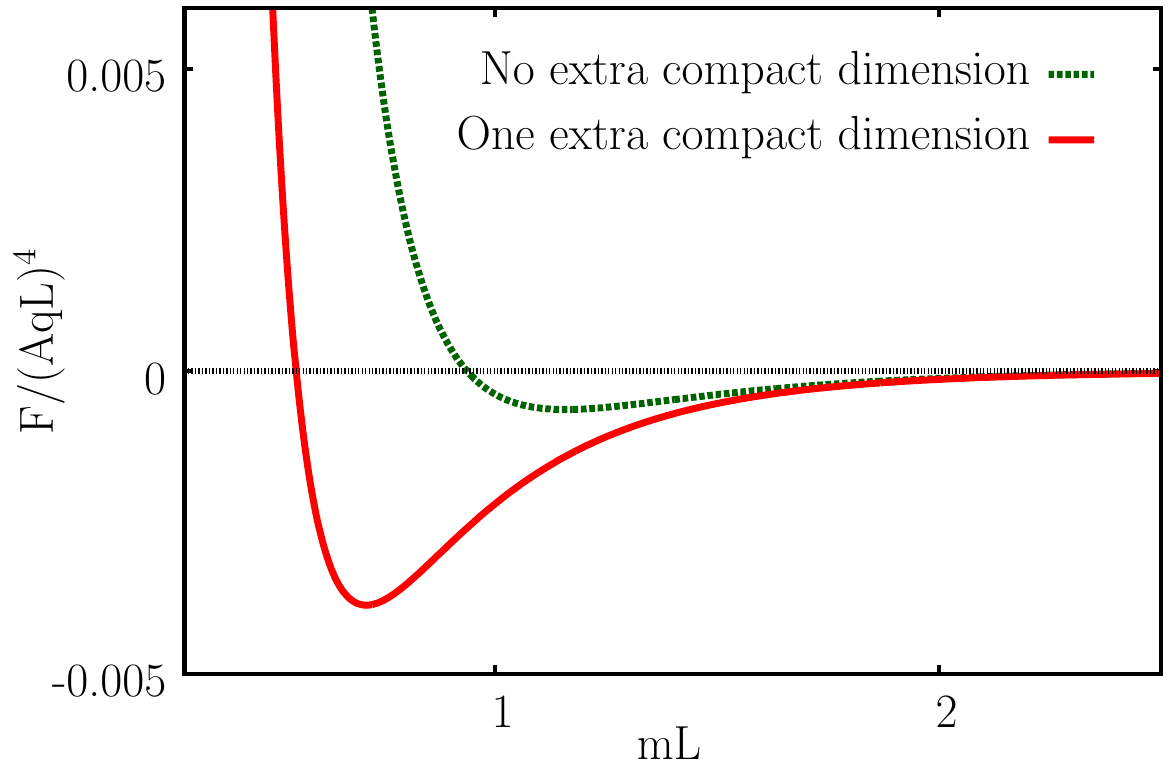}
\caption{Shows the qualitative behaviour of Casimir force in the presence of weak external field and extra compact dimensions. On the y-axis $A$ can be electric field $E$ parallel to the plates or magnetic field $B$ perpendicular to the plates. In this plot we assume that the size of extra compact dimension is same as the seperation between the plates.}
\label{forceb}
\end{figure}

Consider a  complex scalar field in $1+3$ spacetime where one of the spatial dimension is compact under the identification $x_{1}\sim x_{1}+L$. We choose the guage potential $A_{\mu}=(0,a,0,Bx_{2})$ such that there exist a constant magnetic field $B$ along the compact dimension along with a constant guage potential $a$. One can solve Eq. (\ref{heateqn}) with Eq. (\ref{bc1}) by expanding $K(s;x,x')$ in Fourier modes. Heat kernel in the coinciding limit ($x\rightarrow x'$) is
\beq
K(s;x,x)&& = \frac{qB}{L}\l \frac{q B}{2\pi \sinh(2qBs)}\r^{\frac{1}{2}}\times\\
&&\sum_{n=-\infty}^{\infty}\int \frac{d\omega}{2\pi}\frac{dy}{2\pi}e^{-\omega^{2}s}\exp\l -\tanh(qBs)qB y^{2}\r e^{-M_{n}^{2}s}\nn,
\eeq
where $M_{n}^{2}=m^{2}+\frac{4\pi^{2}}{L^{2}}\l n-\frac{q aL}{2\pi}\r^{2}$. Using standard Gaussian integral
\beq\label{kernalb}
K(s;x,x)\= \frac{qB}{8\pi^{\frac{3}{2}}L}\frac{1}{\sqrt{s}\sinh(qBs)}\sum_{n=-\infty}^{\infty}e^{-M_{n}^{2}s}.
\eeq
Substituting Eq. (\ref{kernalb}) in Eq. (\ref{eupot}) gives the Euclidean effective potential as
\beq\label{veb}
V_{E}\= \frac{1}{8\pi^{3/2}L}\sum_{n=-\infty}^{\infty}\int_{0}^{\infty}\frac{ds}{s^{5/2}}\frac{qBs}{\sinh(qBs)}e^{-M_{n}^{2}s}.
\eeq
For a weak magnetic field the renormalized (see appendix \ref{appren}) Euclidean effective potential is
\beq\label{vebren}
V_{E ren}\approx \frac{7q^{4}B^{4}}{2880\pi^{3/2}L}\sum_{n}\int_{0}^{\infty}ds s^{3/2}e^{-M_{n}^{2}s},
\eeq
where we neglect the higher order terms in $B$. Using Gamma function
\beq
V_{Eren}\=\frac{7 q^{4}B^{4}}{3840 L\pi}\l \frac{L}{2\pi}\r^{5}\sum_{n=-\infty}^{\infty}\l \frac{m^{2}L^{2}}{4\pi^{2}}+\l n-\frac{q a L}{2\pi}\r^{2}\r^{-\frac{5}{2}}.
\eeq
Using Eq. (\ref{sum}) we can do the summation which leads to,
\beq
V_{Eren}\= \frac{7 B^{4}q^{4}}{5760 m^{4}\pi^{2}} + \frac{7B^{4}L^{4}q^{4}}{30720\pi^{6}} f_{\lambda}(\alpha,\beta),
\eeq
where $\alpha=m L_{1}/2\pi$ , $\beta=-qaL_{1}/2\pi$ and $\lambda=5/2$. The first term in $V_{Eren}$ is the contribution from $1+3$ flat spacetime \cite{paddy}.  Similar to the above section \ref{seccon}, we choose the constant gauge potential ($a$),  such that it minimizes the Lorentzian effective potential. This gives the vacuum energy of a complex scalar field as
\beq\label{bmin}
V_{min}\= -\frac{7 B^{4}q^{4}}{5760 m^{4}\pi^{2}}\l 1+m^{2}L^{2}\sum_{n=1}^{\infty}n^{2}K_{2}(\xi/2)\r,
\eeq
where we used Eq. (\ref{bessel}).  Including the higher order magnetic field contribution we can rewrite the vacuum energy as
\beq
 V_{min}\= \sum_{k=2}^{\infty}\frac{(2^{2k}-2)}{16\pi^{2}(2k)!}m^{4-4k}B^{2k}q^{2k}B_{2k}\Gamma(2k-2)\\
 &&+ \sum_{k=2}^{\infty}\frac{(1-2^{1-2k})}{\pi^{2}(2k)!}\l\frac{L}{m}\r^{2k-2}B^{2k}q^{2k}B_{2k}\sum_{n=1}^{\infty}n^{2k-2}K_{2k-2}(\xi/2),\nn
 \eeq
where $B_{k}$ is the $(k)^{th}$ Bernoulli number. This matches with the results in \cite{pintomass}. First part of the equation is the contribution from flat space topology \cite{heisenberg}. Now we calculate the Casimir force acting on piston as discussed in section \ref{secforce}. Substituting Eq.  (\ref{bmin}) in Eq. (\ref{rho}) and using Eq. (\ref{fforce}) , the Casimir force per unit area on the piston is
\beq\label{bnforce}
F \= \frac{7 B^{4}q^{4}L^{2}}{5760 m^{4}\pi^{2}}\sum_{n=1}^{\infty}\l 3\xi^{2}K_{2}(\xi) -\frac{1}{2} \xi^{3}\left[ K_{1}(\xi)+K_{3}(\xi)\right]\r.
\eeq
where $\xi=2mnL$. Similarly, we calculate the Casimir force per unit area on the piston with extra compact spatial dimensions, and the results are as shown in the Fig. \ref{forceb}. 

\section{Electric field parallel to the piston and the Schwinger effect}\label{sece}
Consider the guage potential $A_{\mu}=(0,a,0,E\tau)$, a constant guage potential along the compact dimension and a constant electric field along the non compact dimension. Proceeding the same way as discussed in the previous section \ref{secb},  the effective potential is 
\beq\label{effve}
V\=-\frac{1}{8\pi^{3/2}L_{1}}\sum_{n=-\infty}^{\infty}\int_{0}^{\infty}\frac{ds}{s^{5/2}}\frac{qEs}{\sin(qEs)}e^{-M_{n}^{2}s}.
\eeq
This explicitly shows the eletric-magnetic duality in the effective potential \cite{duality1,duality2}. Following the calculations in appendix \ref{appren} and section \ref{secb},  for a weak electric field $E\rightarrow 0$, one can calculate the Casimir force per unit area on the piston as
\beq\label{enforce}
F \= \frac{7 E^{4}q^{4}}{5760 m^{4}\pi^{2}}\sum_{n=1}^{\infty}\l 3 \xi^{2}K_{2}(\xi) -\frac{1}{2} \xi^{3}\left[ K_{1}(\xi)+K_{3}(\xi) \right]\r.
\eeq
Similar calculations can be extended to $d$ dimensional spacetime with $k$ compact spatial dimensions.  For the case of one extra compact dimension, the corresponding Casimir force is
\beq\label{e1comf}
F = -\frac{7E^{4}q^{4}}{11520\pi^{2}m^{4}}&&\sum_{n=1}^{\infty}\Big( \xi^{2} K_{2}(\xi)+\frac{Rm}{\sqrt{2\pi}}\xi^{\frac{3}{2}} K_{\frac{3}{2}}(\xi)\nn\\
&&- \xi^{3}K_{1}(\xi) -\frac{Rm}{\sqrt{2\pi}}\xi^{\frac{5}{2}}K_{\frac{1}{2}}(\xi)\Big),
\eeq
where $\xi = 2mnL$ as usual and $R$ is the size of extra compact dimension. A plot showing the influence of extra compact dimension on the Casimir force is given in Fig. \ref{forceb}.  In the presence of electric field $E$, the effective potential has an imaginary contribution coming  from the poles at $s_{k}= k\pi/qE$ where $k\in N$ (Eq.  \ref{effve}). The imaginary part of the effective potential is (see appendix \ref{appcont})
\beq\label{schwing}
Im(V)\= \sum_{k=1}^{\infty}\frac{(-1)^{k+1}}{2(2\pi)^{3}}e^{\frac{-k\pi m^{2}}{qE}}\l\frac{qE}{k}\r^{2}\Theta\l 3,0,e^{\frac{-q EL^{2}}{4k\pi}}\r,
\eeq
where $\Theta$ is the elliptic theta function.  Here we  used $M_{n}^{2}=m^{2}+\frac{4\pi^{2}}{L^{2}}\l n-\frac{q aL}{2\pi}\r^{2}$ and also  took $ a=0$ as that choice minimizes the potential.  The imaginary part of the effective potential gives the number of pairs of charged scalar particles created by the electric field per unit time per unit volume.  In the limit $L\rightarrow\infty$, $\Theta\rightarrow 1$ and gives the standard Schwinger effect result in $3+1$ dimension \cite{kim,schwinger}. 
\begin{figure}[H]
\centering
\includegraphics[scale=0.6]{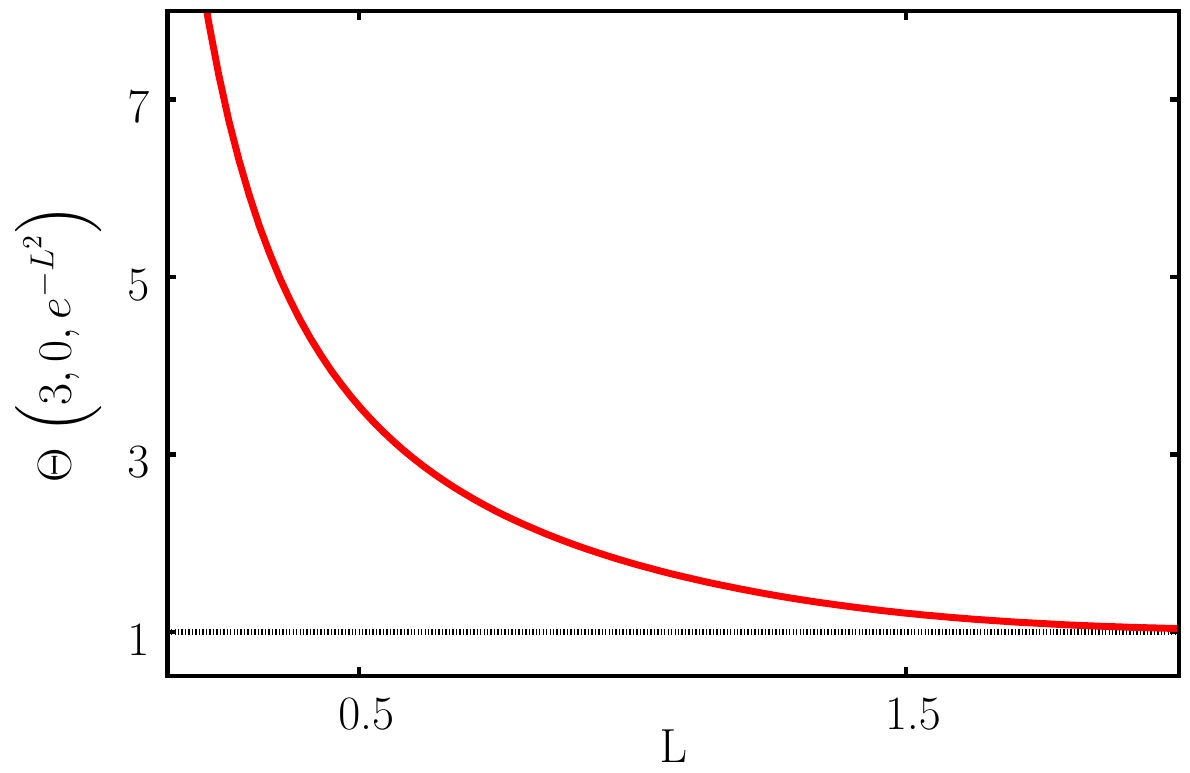}
\caption{Shows the variation of imaginary part of the effective potential with respect to the separation between the parallel plates.}
\label{figschwinger}
\end{figure}
Presence of extra compact dimensions, along with the piston, add the number of $\Theta$ functions corresponding to each extra compact dimensions. But if we consider extra compact dimensions without the piston, i.e.,  $A_{\mu}=(0,0,0,E\tau,a)$. The corresponding effective potential is
\beq
V = -\frac{q E}{16\pi^{2}R}\sum_{n=-\infty}^{\infty}\int_{0}^{\infty}\frac{ds}{s^{2}}\frac{e^{-M_{n}^{2}}s}{\sin(q E s)},
\eeq
where $R$ is the size of extra compact dimension and $M_{n}^{2}= (\frac{2\pi n}{R}-qa)^{2}+m^{2}$.  There are no branch cuts in the integral and using the results in \cite{kim,paddy} 
\beq\label{schwingcom}
Im(V)\=  \sum_{k=1}^{\infty}\frac{(-1)^{k+1}}{4(2\pi)^{3}}e^{\frac{-k\pi m^{2}}{qE}}\l\frac{qE}{k}\r^{\frac{5}{2}}\Theta\l 3,0,e^{\frac{-q ER^{2}}{4k\pi}}\r.
\eeq
The limit $R\rightarrow \infty$ gives the Schwinger effect in $1+4$ flat dimension. In the same way, one can calculate the pair production in a higher compact and non-compact dimensions, with or without pistons. 

\section{Summary and conclusion}\label{summary}
In short, Eq. \ref{apprx}, \ref{bnforce}, \ref{e1comf}, \ref{schwing} and \ref{schwingcom} are  the central results of this paper, hitherto unreported in previous literature.

Summarizing, this paper considers the influence of extra compact dimensions and external background gauge fields on the Casimir force on a piston using effective action method. We discuss the influence of both magnetic fields perpendicular to the piston, as well as the electric field parallel to the piston in the presence of extra compact dimensions. In the case of an electric field, we also considered the effect of parallel plates and extra compact dimensions on the Schwinger particle production process.

In section \ref{secforce}, we examine the effect of extra compact dimensions on Casimir forces. A similar study for the massless scalar field using the mode sum approach is presented in \citep{cheng2}. For massive scalar fields in lower and non-compact dimensions, our results match with the conclusions in \cite{pinto,massd,massd2}. The Casimir force acting between the parallel plates is always attractive irrespective of the mass, the number of extra compact dimensions, and the size of extra compact dimensions (Figure \ref{figmass}). However, the presence of extra compact dimensions decrease the force of attraction between the parallel plates.  This may be because of the following reason. In the presence of extra dimensions, the vacuum fluctuations can escape into these dimensions, resulting in decreased modes in the presence of a piston. This hypothesis is supported by the observation that the Casimir force per unit area per unit length of extra compact dimension is maximum when the size of the extra compact dimension is less than the separation between the plates.The influence of the extra compact dimensions is maximum in the regime $Lm< 1$ (Figure \ref{figmass}). Casimir force is maximal for a massless scalar field and decreases as mass increases. Casimir force vanishes to zero as $m\rightarrow \infty$ as there are no more quantum fluctuations in this limit. Corrections in Casimir force per unit area on the piston is more significant for large compact dimensions. Similar conclusions can be obtained for massless scalar fields.

Experimentally, the Casimir effect has been observed for a parallel plate separation of the order $10^{-7}$m \cite{PhysRevLett.78.5,Bordag:2009zz}. For the Higgs field (mass $\approx 10^{-25}$Kg \cite{2020135425}), using the above values, the Casimir force per unit area on the piston without any extra compact dimension is $-4.318\times 10^{-8}N$. Now from phenomenological models, if we consider the upper bound on the size of one extra compact dimension as $R \leq 10$ nm \cite{poppenhaeger2004casimir}, we see from our Eq.  (\ref{schwing}) that the presence of one extra compact dimension almost halves the intensity of attraction to $-2.262\times 10^{-8}N$. These qualitative numbers can in fact provide us with a test bed to probe for existence of compact extra dimensions. 

Additionally, in sections \ref{secb} and \ref{sece}, we explore the influence of external background fields on the Casimir force in the presence of extra compact dimensions. If we take the limit $B$ to infinity in Eq.  (\ref{veb}), the effective potential vanishes to zero. From this, we conclude that a high magnetic field always inhibits the Casimir force as mentioned in \citep{pintomass}. But in the presence of a high electric field parallel to the piston, the effective potential is instead enhanced  as seen from Eq.  \ref{effve}. However, in the weak field approximation, both electric and magnetic field enhance the Casimir force of attraction as observed from  Eqs.  (\ref{enforce} and \ref{bnforce}). Also, in the weak field approximation, there is a repulsive contribution to the Casimir force for $Lm<1$ (Figure \ref{forceb}). For the case of a weak magnetic field, this repulsive contribution is noted in \cite{pallab}. Figure \ref{forceb} illustrates that even though the extra compact dimension does not change the qualitative behavior of Casimir force, it does decrease its magnitude. The electric field can also produced charged pairs of particle from the vacuum, called the Schwinger effect. The Schwinger effect still awaits experimental confirmation in the laboratory as it requires extreme electric fields. Our results \ref{schwing} show that the presence of parallel plates can enhance the pair production. This observation can help in the experimental realization of the Schwinger effect. Additionally, we can deduce from \ref{schwingcom} that the presence of extra compact dimensions also influences the Schwinger pair product.

The present formalism can be further extended for studying twisted fields in similar manifolds, and also for investigating Casimir effect at finite temperatures which will be discussed in later investigations.

\newpage
\appendix
\section{Behaviour of $f_{\lambda}(\alpha,\beta)$}\label{appf}
\subsection{The maxima and minima of the function $f_{\lambda}(\alpha,\beta)$ with respect to $\beta$ }\label{beta}\label{B}
We are interested in finding the maximum of $f_{\lambda}(\alpha,\beta)$ with respect to $\beta$ when $\lambda<0$. From Eq. (\ref{sum})
\beq
\frac{\partial f}{\partial \beta} = 2\pi\mathcal{I}\left\{\int_{\alpha}^{\infty}dx\frac{(x^{2}-\alpha^{2})^{-\lambda}e^{2\pi x+2\pi i \beta}}{\left(e^{2\pi x+2\pi i\beta}-1\right)^{2}}\right\}\; ,
\eeq
 where $\mathcal{I}$ gives the imaginary part of the function inside. But,
\beq
\frac{e^{2\pi x+2\pi i\beta}}{\left(e^{2\pi x+2i\pi\beta}-1\right)^{2}} =\Big(&&\l 2\cos(2\pi\beta)\cosh(2\pi x)-2\right)\nn\\
&&+2i\sin(2\pi\beta)\sinh(2\pi x)\Big)^{-1} .
\eeq
Which gives,
\beq
\frac{\partial f}{\partial\beta} \= -4\pi \int_{\alpha}^{\infty} dx \left(x^{2}-\alpha^{2}\right)^{-\lambda}\times\nn\\
&&\frac{\sin(2\pi\beta)\sinh(2\pi x)}{\left(2\cos(2\pi\beta)\cosh(2\pi x)-2\right)^{2}+\left(2\sin(2\pi\beta)\sinh(2\pi x)\right)^{2}}\nn.
\eeq
Then,
\begin{equation}
\begin{split}
\frac{\partial f}{\partial\beta} = 0  \hspace{0.5cm} \forall\hspace{0.5cm}  \beta =\frac{n}{2} \hspace{0.5cm} \mbox{where}\hspace{0.25cm} n \in \mathbf{Z} .\\
\end{split}
\end{equation}
Checking the second derivative at these critical points,
\beq
\frac{\partial^{2} f}{\partial \beta^{2}}\Big|_{\beta=\frac{n}{2}} \=4\pi \int_{\alpha}^{\infty} dx \left(x^{2}-\alpha^{2}\right)^{-\lambda}\times\nn\\
&&\frac{2\pi(-1)^{n+1}}{\left(2\cos(2\pi\beta)\cosh(2\pi x)-2\right)^{2}+\left(2\sin(2\pi\beta)\sinh(2\pi x)\right)^{2}}\; \nn.
\eeq
Which gives
\begin{equation*}
\mbox{Sign}\left(\frac{\partial^{2} f}{\partial \beta^{2}}\right) = (-1)^{n+1} =(-1)^{2\beta+1}\hspace{0.5cm} \forall \hspace{0.25cm}n \in \mathbf{Z}.
\end{equation*}
From this we can conclude that $f$ is maximum when  $\beta = n$ and is minimum when $\beta= n+\frac{1}{2}$. \\
\subsection{Convergence of $f_{\lambda}(\alpha,\beta)$}
\begin{figure}[H]
\centering
\includegraphics[scale=0.6]{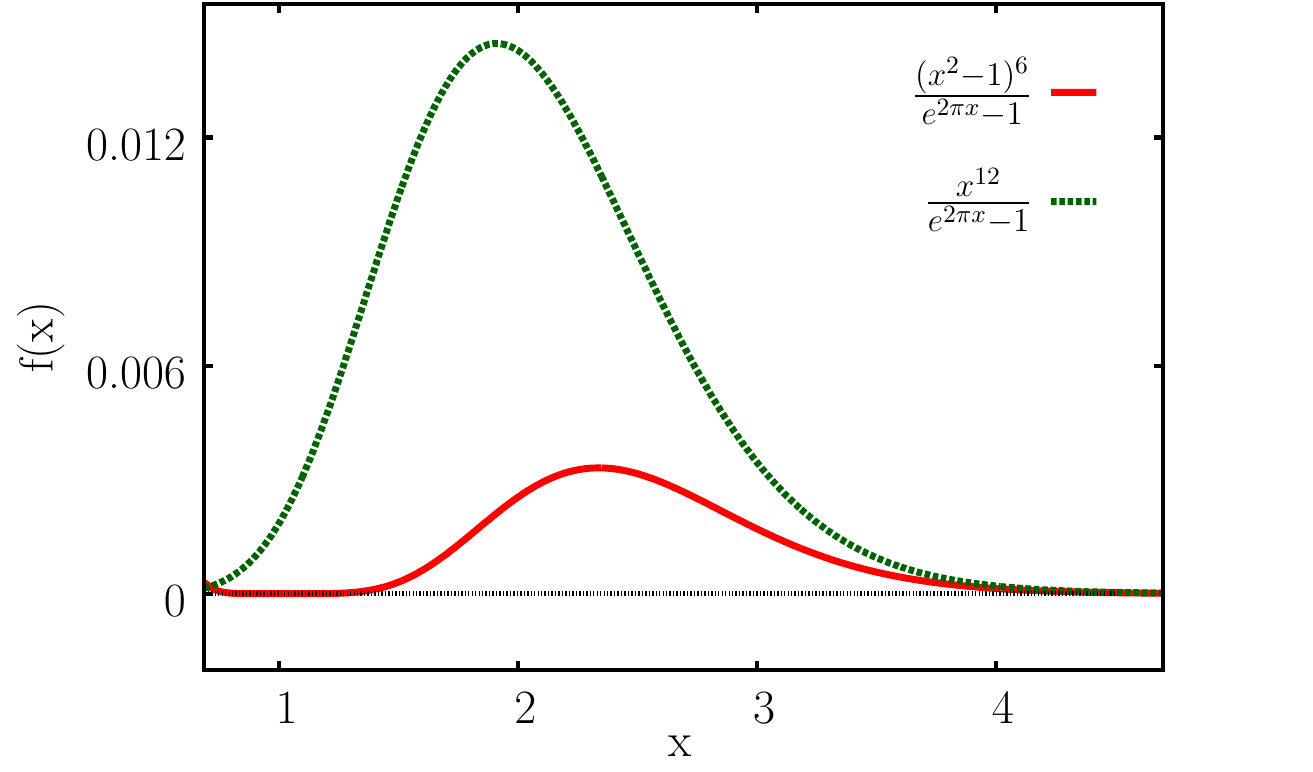}
\caption{Graphical representation of (\ref{con}) for $\alpha=1$ and $\lambda=6$.}
\label{Fig3}
\end{figure}

From appendix \ref{B}, the maximum of $f_{\lambda}(\alpha,\beta)$ with respect to $\beta$ is given by
\begin{equation}
f_{-\lambda_{max}}(\alpha) =\int_{\alpha}^{\infty}dx \frac{\left(x^{2}-\alpha^{2}\right)^{\lambda}}{e^{2\pi x}-1}\; ,
\end{equation}
with $\lambda ,\alpha> 0$. By inspection and from Fig.  \ref{Fig3}  below we can see that,
\begin{equation}\label{con}
\frac{x^{2\lambda}}{e^{2\pi x}-1}\ge \frac{\left(x^{2}-\alpha^{2}\right)^{\lambda}}{e^{2\pi x}-1}.
\end{equation}
But
\begin{equation*}
\int_{0}^{\infty} dx \frac{x^{\lambda}}{e^{2\pi x}-1} = \frac{1}{(2\pi)^{\lambda+1}}\Gamma\left(1+\lambda\right)\mbox{PolyLog}\left(1+\lambda,1\right).
\end{equation*}
Then using comparison test for convergence, we can conclude that $f_{\lambda_{max}}(\alpha)$ is converging, from which we can conclude that (\ref{sum}) is converging.\\
\subsection{Behaviour of the function $f_{\lambda}(\alpha,\beta)$ with respect to  $\alpha$}
\begin{figure}[H]
\centering
\includegraphics[scale=0.6]{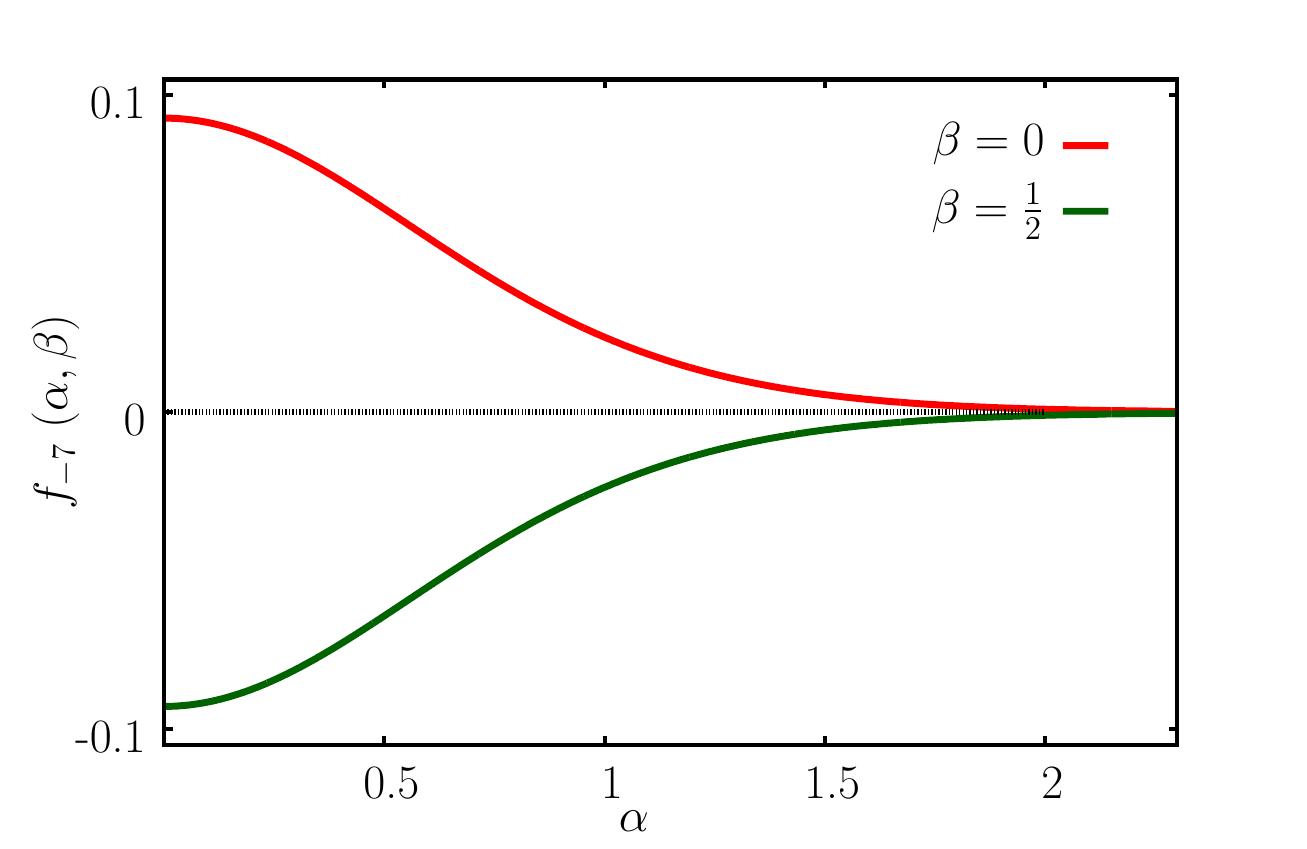}
\caption{Shows the variation of $f_{\lambda}(\alpha,\beta)$ with respect to $\alpha$ for different values of $\lambda$ ($\lambda=-7,-8,-9$ and $ -10$ in plots A, B, C, and D respectively). In each plot we choose $\beta$ such that $\cos(2\pi\beta)>0$ for one and $\cos(2\pi\beta)<0$ for other.}
\label{Fig4}
\end{figure}

From Eq.  (\ref{sum})
\begin{equation*}\label{f}
\begin{split}
f_{-\lambda}(\alpha,\beta) = \int_{\alpha}^{\infty}dx\frac{\left(x^{2}-\alpha^{2}\right)^{\lambda}\left(e^{2\pi x}\cos(2\pi\beta)-1\right)}{\left(e^{2\pi x}\cos(2\pi\beta)-1\right)^{2}+e^{4\pi x}\sin^{2}(2\pi\beta)}.
\end{split}
\end{equation*}
Which gives,
\beq
\frac{\partial f_{-\lambda}(\alpha)}{\partial\alpha} \= -2\alpha\lambda \int_{\alpha}^{\infty} dx\times \\
&&\frac{\left(x^{2}-\alpha^{2}\right)^{\lambda-1}\left(e^{2\pi x}\cos(2\pi\beta)-1\right)}{\left(e^{2\pi x}\cos(2\pi\beta)-1\right)^{2}+e^{4\pi x}\sin^{2}(2\pi\beta)}\nn.
\eeq
In our case interest $\lambda\geq 1$(which is half the spacetime dimensions) and $\alpha\ge 0$. So $f_{\lambda}(\alpha,\beta)$ is a monotonically decreasing function if $\cos(2\pi\beta)>0$ and is monotonically increasing function if $\cos(2\pi\beta)<0$. But in both cases the magnitude of the integral will be maximum when $\alpha=0$. From the plots below (Fig. \ref{Fig4}) we can see that the $\sum_{n_2}f_{\lambda}\left(\alpha_{n_{1}n_{2}},\beta\right)$  is maximum when $\cos(2\pi\beta)>0$, which gives the positive value for each term in the sum and $\beta\in \mathbb{Z}$. Also the fast decay of the function shows that the major contribution comes when $\alpha = 0$. For higher summations also this factor remains the same, which is checked numerically.
\section{Renormalization}\label{appren}
The Euclidean effective potential is
\beq
V_{E}\= \int_{0}^{\infty}\frac{ds}{s}K(s;x,x)\nn\\
\= \frac{1}{8\pi^{3/2}L_{1}}\sum_{n=-\infty}^{\infty}\int_{0}^{\infty}\frac{ds}{s^{5/2}}\frac{qBs}{\sinh(qBs)}e^{-M_{n}^{2}s}.
\eeq
In the limit $B\rightarrow 0$
\beq
V_{E}|_{B\rightarrow 0}\= \frac{1}{8\pi^{3/2}L_{1}}\sum_{n=-\infty}^{\infty}\int_{0}^{\infty}\frac{ds}{s^{5/2}}e^{-M_{n}^{2}s}.
\eeq
Subtracting this will leads to the new effective potential as
\beq
V1_{E}\=\frac{1}{8\pi^{3/2}L_{1}}\sum_{n=-\infty}^{\infty}\int_{0}^{\infty}\frac{ds}{s^{5/2}}e^{-M_{n}^{2}s}\l \frac{qBs}{\sinh(qBs)}-1\r\nn.
\eeq
Near $s=0$ the effective potential is given by
\beq
V_{1}\approx&& \frac{1}{8\pi^{3/2}L_{1}}\sum_{n}\frac{B^{2}q^{2}}{6}\int_{0}^{\infty}\frac{ds}{\sqrt{s}}e^{-M_{n}^{2}s}\nn\\
&&-\frac{7}{360}q^{4}B^{4}\int_{0}^{\infty}ds s^{3/2}e^{-M_{n}^{2}s}+O(B^{6}),
\eeq
where we took $V=-V_{E}$. The first term is proportional to $B^{2}$ which corresponds to the classical Lagrangian. The total Lagrangian for the system is
\beq
L_{tot}\= L_{0}+V = L_{0}+L_{c}+\l V-L_{c}\r,
\eeq
where
\beq
L_{c}\= \frac{-1}{8\pi^{3/2}L_{1}}\sum_{n}\frac{B^{2}q^{2}}{6}\int_{0}^{\infty}\frac{ds}{\sqrt{s}}e^{-M_{n}^{2}s}\equiv B^{2}Z.
\eeq
Now subtracting out $L_{c}$ from $V_{1}$ and redefining the field strength and charges as
\beq
B'\= (1+Z)^{\frac{1}{2}}B\;\; ; q' = (1+Z)^{-\frac{1}{2}}q.
\eeq
Then the renormalized effective potential is 
\beq
V_{ren}\=-\frac{1}{8\pi^{3/2}L_{1}}\sum_{n}\frac{7}{360}q^{4}B^{4}\int_{0}^{\infty}ds s^{3/2}e^{-M_{n}^{2}s}\\
&&+O(B^{6})\nn,
\eeq
where we wrote $ B$ and $q$ for $B'$ and $q'$.
\section{Contour integral}\label{appcont}
Consider
\beq\label{eqn1}
I = \int_{0}^{\infty}\frac{ds}{s^{5/2}}\frac{qEs}{\sin(qEs)}e^{-M_{n}^{2}s}
\eeq
This has branch points at $s=0$ and $s=\infty$. Also has singularities at $s_{k}=\frac{k\pi}{q E}$ where $k\in N$.  But the singularity at $s=0$ is already considered for calculating Casimir energy or the real part of the effective potential. Then the contribution to the imaginary part of the effective potential is coming from other poles.
\begin{figure}[H]\label{contour}
\centering
\includegraphics[scale=0.7]{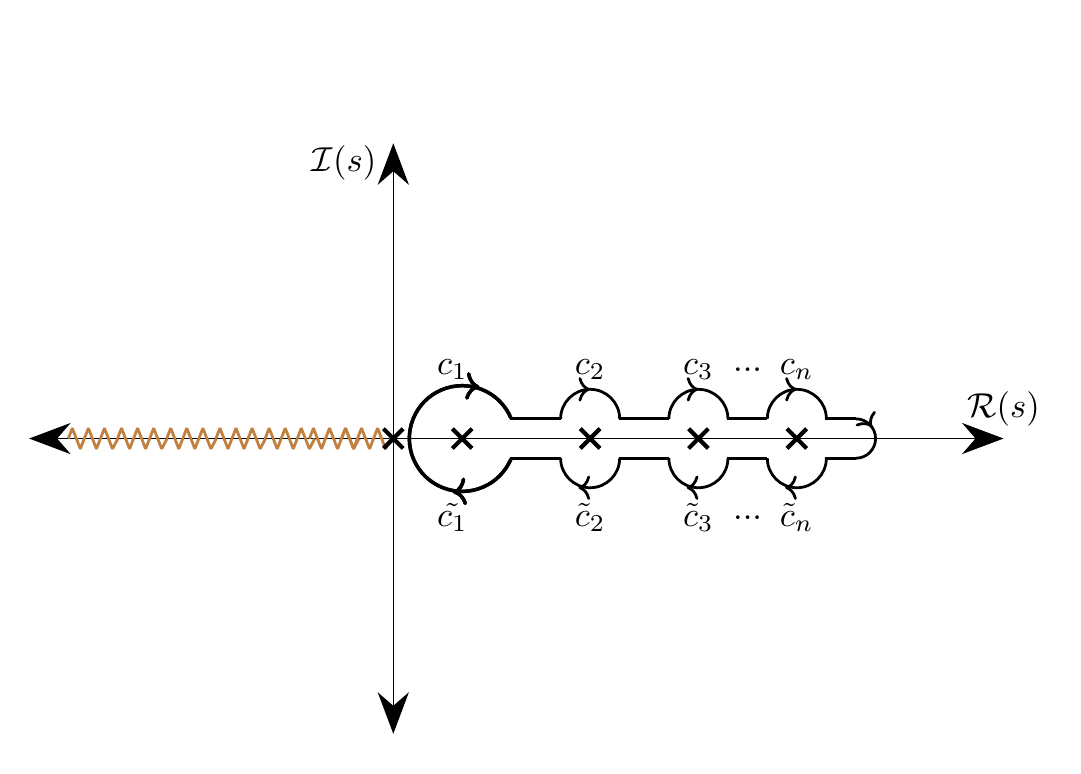}
\caption{The contour of s integration for Eq.  \ref{eqn1}}
\end{figure}
Taking
\beq
f(z)\= \frac{1}{z^{5/2}}\frac{qEz}{\sin(qEz)}e^{-M_{n}^{2}z},
\eeq
and using Cauchy's theorem we can say that
\beq
\oint f(z)dz =2\pi i \sum_{k=1}^{\infty}(-1)^{k}\l\frac{qE}{k\pi}\r^{\frac{3}{2}}e^{-\frac{M_{n}^{2}k\pi}{q E}},
\eeq
along the closed contour given in the Fig.  \ref{contour}.  From this we can write
\beq
\oint f(z)dz \=\int_{c_{1}}f(z)dz+ \int_{s_{1}+\epsilon}^{s_{2}-\epsilon}f(z)dz+...\\
&&+\int_{-\infty}^{s_{n}+\epsilon}+\int_{\tilde{c_{n}}}f(z)dz+...\int_{s_{1}-\epsilon}^{\epsilon}f(z)dz\nn.
\eeq
In the fourth quadrant or in the lower half plane the function will have an extra $2\pi$ phase which gives
\beq
f(z)\=\frac{e^{-i\pi}}{z^{5/2}}\frac{qEz}{\sin(qEz)}e^{-M_{n}^{2}z}= -f(z).
\eeq  
Then in the limit $\epsilon\rightarrow 0$ this gives
\beq
\oint f(z)dz\= 2I + \int_{c_{1}}f(z)dz+\int_{c_{2}}f(z)dz + ...\\
&&+\int_{c_{n}}f(z)dz+\int_{\tilde{c_{n}}}f(z)+...\int_{\tilde{c_{1}}}f(z)dz.\nn
\eeq
Then for integration over $c_{1}$, take $z=s_{1}+\epsilon e^{i\theta}$
\beq
\int_{c_{1}}f(z)dz\=\int_{\pi}^{0}  \frac{i\epsilon e^{i\theta}d\theta}{s_{1}^{5/2}}\frac{qEs_{1}}{\cos(qEs_{1})\epsilon e^{i\theta}}e^{-M_{n}^{2}s_{1}},\nn\\
\= -i\frac{\pi q E e^{-M_{n}^{2}s_{1}}}{s_{1}^{\frac{3}{2}}\cos(qE s_{1})}.
\eeq
The integration over $c_{1}$ and $\tilde{c_{1}}$ add up to zero, similarly for all other $c_{n}'s$.
From this we can write
\beq
I =\pi i \sum_{k=1}^{\infty}(-1)^{k}\l\frac{qE}{k\pi}\r^{\frac{3}{2}}e^{-\frac{M_{n}^{2}k\pi}{q E}},
\eeq
\section{Mode sum approach}
The infinite sum in Eq. (\ref{sum}) leads to
\beq\label{2}
\sum_{n=-\infty}^{\infty}\left(n^{2}+\alpha^{2}\right)^{-\lambda} \= \frac{\sqrt{\pi}\Gamma\left(\lambda-\frac{1}{2}\right)\alpha^{1-2\lambda}}{\Gamma(\lambda)}\\
&&+4\sin(\lambda\pi)f_{\lambda}(\alpha,0),\nn
\eeq
and
 \beq\label{3}
  \sum_{n=1}^{\infty}\left(n^{2}+\alpha^{2}\right)^{-\lambda} \=-\frac{1}{2}\alpha^{-2\lambda}+ \frac{\sqrt{\pi}\Gamma\left(\lambda-\frac{1}{2}\right)\alpha^{1-2\lambda}}{2\Gamma(\lambda)}\\
  &&+2\sin(\lambda\pi)f_{\lambda}(\alpha,0).\nn
 \eeq
The energy of the complex scalar field with one extra compact dimension and having two plates separated by a distance of $L$ is given by
 \begin{equation}
 \begin{split}
 E &= \int \frac{d^{2}k}{(2\pi)^{2}}\sum_{n=1}^{\infty}\sum_{n_{1}=-\infty}^{\infty}\sqrt{k_{1}^{2}+k_{2}^{2}+\frac{n^{2}\pi^{2}}{L^{2}}+\frac{4n_{1}^{2}\pi^{2}}{L_{1}^{2}}}\; ,\\
 &=\frac{B\left(1,-\frac{3}{2}\right)}{4\pi}\sum_{n=1}^{\infty}\sum_{n_{1}=-\infty}^{\infty}\left(\frac{n^{2}\pi^{2}}{L^{2}}+\frac{4n_{1}^{2}\pi^{2}}{L_{1}^{2}}\right)^{\frac{3}{2}}\; ,
 \end{split}
 \end{equation} 
where we used the integral representation of Beta function. Considering the summation in both order leads to
\beq
  E =\frac{-1}{12\pi}\Bigg(\sum_{n=1}^{\infty}\sum_{n_{1}=-\infty}^{\infty}&&\left(\frac{n^{2}\pi^{2}}{L^{2}}+\frac{4n_{1}^{2}\pi^{2}}{L_{1}^{2}}\right)^{\frac{3}{2}}\\
  &&+\sum_{n_{1}=-\infty}^{\infty}\sum_{n=1}^{\infty}\left(\frac{n^{2}\pi^{2}}{L^{2}}+\frac{4n_{1}^{2}\pi^{2}}{L_{1}^{2}}\right)^{\frac{3}{2}}\Bigg).\nn
\eeq
Using Eq. (\ref{2}) and Eq. (\ref{3}) 
\begin{eqnarray}
 \sum_{n=1}^{\infty}\sum_{n_{1}=-\infty}^{\infty}\left(\frac{n^{2}\pi^{2}}{L^{2}}+\frac{4n_{1}^{2}\pi^{2}}{L_{1}^{2}}\right)^{\frac{3}{2}}&& = \frac{9L_{1}\zeta(5)}{32\pi L^{4}}\nn\\
 &&+\frac{32\pi^{3}}{L_{1}^{3}}\sum_{n=1}^{\infty}f_{-\frac{3}{2}}\left(\frac{L_{1}n}{2L},0\right),\nn\\
\sum_{n_{1}=-\infty}^{\infty}\sum_{n=1}^{\infty}\left(\frac{n^{2}\pi^{2}}{L^{2}}+\frac{4n_{1}^{2}\pi^{2}}{L_{1}^{2}}\right)^{\frac{3}{2}}  &=& \frac{\pi^{3}}{120}\left(\frac{1}{L^{3}}-\frac{8}{L_{1}^{3}}\right)\nn\\
&&+\frac{9L\zeta(5)}{\pi L_{1}^{4}}\\
&&+\frac{4\pi^{3}}{L^{3}}\sum_{n=1}^{\infty}f_{-\frac{3}{2}}\left(\frac{2Ln}{L_{1}},\right).\nonumber
\end{eqnarray}
where we have used $\Gamma(-2)\zeta(-4)=\frac{3\zeta(5)}{4\pi^{4}}$. This gives,
\begin{eqnarray}
  E&=&\frac{\pi^{2}}{180}\left[\frac{1}{L_{1}^{3}}-\frac{1}{8L^{3}}\right]-\frac{3L\zeta(5)}{4\pi^{2}L_{1}^{4}}-\frac{3L_{1}\zeta(5)}{128\pi^{2}L^{4}}\\
  &-&\frac{\pi^{2}}{3L^{3}}\sum_{n=1}^{\infty}f_{-\frac{3}{2}}\left(\frac{2Ln}{L_{1}},0\right)-\frac{8\pi^{2}}{3L_{1}^{3}}\sum_{n=1}^{\infty}f_{-\frac{3}{2}}\left(\frac{L_{1}n}{2L},0\right).\nonumber
\end{eqnarray}
Now considering the system in a finite box as in Fig.  \ref{piston}, the energy in region I is given by
\begin{eqnarray}
E(L) &=& \frac{\pi^{2}}{180}\left[\frac{1}{L_{1}^{3}}-\frac{1}{8L^{3}}\right]-\frac{3\zeta(5)L_{1}}{128\pi^{2}L^{4}}-\frac{3\zeta(5)L}{4\pi^{2}L_{1}^{4}}\\
&-&\frac{8\pi^{2}}{3L_{1}^{2}}\sum_{n=1}^{\infty}f_{-\frac{3}{2}}\left(\frac{nL_{1}}{2L},0\right)-\frac{\pi^{2}}{3L^{3}}\sum_{n_{1}=1}^{\infty}f_{-\frac{3}{2}}\left(\frac{2Ln_{1}}{L_{1}},0\right).\nonumber
\end{eqnarray}
Similarly the energy of region II is given by
\beq
E(l_{1}-L)\= \frac{\pi^{2}}{180}\left[\frac{1}{L_{1}^{3}}-\frac{1}{8(l_{1}-L)^{3}}\right]-\frac{3\zeta(5)L_{1}}{128\pi^{2}(l_{1}-L)^{4}}\nn\\
-&&\frac{3\zeta(5)(l_{1}-L)}{4\pi^{2}L_{1}^{4}}
-\frac{8\pi^{2}}{3L_{1}^{2}}\sum_{n=1}^{\infty}f_{-\frac{3}{2}}\left(\frac{nL_{1}}{2(l_{1}-L)},0\right)\nn\\
&&-\frac{\pi^{2}}{3(l_{1}-L)^{3}}\sum_{n_{1}=1}^{\infty}f_{-\frac{3}{2}}\left(\frac{2(l_{1}-L)n_{1}}{L_{1}},0\right).
\eeq
Now calculating the Casimir force as in \cite{cheng2},
\beq
F\=-\frac{d}{dL} \left( E(L)+E(l_{1}-L)\right)\big|_{l_{1}\rightarrow \infty} \; ,\nn\\
\= -\frac{\pi^{2}}{480 L^{4}}-\frac{3\zeta(5)L_{1}}{32\pi^{2}L^{5}}\nn\\
&& -\frac{\pi^{2}}{L^{4}}\sum_{n=1}^{\infty}f_{-\frac{3}{2}}\left(\frac{2Ln}{L_{1}},0\right)-\frac{4\pi^{2}}{L^{2}L_{1}^{2}}\sum_{n=1}^{\infty}n^{2}f_{-\frac{1}{2}}\left(\frac{2Ln}{L_{1}},0\right)\nn\\
&&+\frac{2\pi^{2}}{L_{1}L^{3}}\sum_{n=1}^{\infty}n^{2}f_{-\frac{1}{2}}\left(\frac{nL_{1}}{2L},0\right).
\eeq
This matches with the result we got before upto the approximation, i.e., the first two terms matches and we neglect other terms in our approximation.

\bibliographystyle{apsrev4-2}
\bibliography{reference}
\end{document}